\PassOptionsToPackage{unicode}{hyperref}
\PassOptionsToPackage{hyphens}{url}
\PassOptionsToPackage{dvipsnames,svgnames,x11names}{xcolor}
\documentclass[
  12pt]{article}

\usepackage{amsmath,amssymb}
\usepackage{iftex}
\ifPDFTeX
  \usepackage[T1]{fontenc}
  \usepackage[utf8]{inputenc}
  \usepackage{textcomp} 
\else 
  \usepackage{unicode-math}
  \defaultfontfeatures{Scale=MatchLowercase}
  \defaultfontfeatures[\rmfamily]{Ligatures=TeX,Scale=1}
\fi
\usepackage{lmodern}
\ifPDFTeX\else  
\fi
\IfFileExists{upquote.sty}{\usepackage{upquote}}{}
\IfFileExists{microtype.sty}{
  \usepackage[]{microtype}
  \UseMicrotypeSet[protrusion]{basicmath} 
}{}
\makeatletter
\@ifundefined{KOMAClassName}{
  \IfFileExists{parskip.sty}{%
    \usepackage{parskip}
  }{
    \setlength{\parindent}{0pt}
    \setlength{\parskip}{6pt plus 2pt minus 1pt}}
}{
  \KOMAoptions{parskip=half}}
\makeatother
\usepackage{xcolor}
\ifx\paragraph\undefined\else
  \let\oldparagraph\paragraph
  \renewcommand{\paragraph}[1]{\oldparagraph{#1}\mbox{}}
\fi
\ifx\subparagraph\undefined\else
  \let\oldsubparagraph\subparagraph
  \renewcommand{\subparagraph}[1]{\oldsubparagraph{#1}\mbox{}}
\fi

\providecommand{\tightlist}{%
  \setlength{\itemsep}{0pt}\setlength{\parskip}{0pt}}\usepackage{longtable,booktabs,array}
\usepackage{calc} 
\usepackage{etoolbox}
\makeatletter
\patchcmd\longtable{\par}{\if@noskipsec\mbox{}\fi\par}{}{}
\makeatother
\IfFileExists{footnotehyper.sty}{\usepackage{footnotehyper}}{\usepackage{footnote}}
\makesavenoteenv{longtable}
\usepackage{graphicx}
\makeatletter
\def\maxwidth{\ifdim\Gin@nat@width>\linewidth\linewidth\else\Gin@nat@width\fi}
\def\maxheight{\ifdim\Gin@nat@height>\textheight\textheight\else\Gin@nat@height\fi}
\makeatother
\setkeys{Gin}{width=\maxwidth,height=\maxheight,keepaspectratio}
\makeatletter
\def\fps@figure{htbp}
\makeatother

\makeatletter
\@ifpackageloaded{caption}{}{\usepackage{caption}}
\AtBeginDocument{%
\ifdefined\contentsname
  \renewcommand*\contentsname{Table of contents}
\else
  \newcommand\contentsname{Table of contents}
\fi
\ifdefined\listfigurename
  \renewcommand*\listfigurename{List of Figures}
\else
  \newcommand\listfigurename{List of Figures}
\fi
\ifdefined\listtablename
  \renewcommand*\listtablename{List of Tables}
\else
  \newcommand\listtablename{List of Tables}
\fi
\ifdefined\figurename
  \renewcommand*\figurename{Figure}
\else
  \newcommand\figurename{Figure}
\fi
\ifdefined\tablename
  \renewcommand*\tablename{Table}
\else
  \newcommand\tablename{Table}
\fi
}
\@ifpackageloaded{float}{}{\usepackage{float}}
\floatstyle{ruled}
\@ifundefined{c@chapter}{\newfloat{codelisting}{h}{lop}}{\newfloat{codelisting}{h}{lop}[chapter]}
\floatname{codelisting}{Listing}

\makeatother
\makeatletter
\@ifpackageloaded{caption}{}{\usepackage{caption}}
\@ifpackageloaded{subcaption}{}{\usepackage{subcaption}}
\makeatother
\makeatletter
\@ifpackageloaded{tcolorbox}{}{\usepackage[skins,breakable]{tcolorbox}}
\makeatother
\makeatletter
\@ifundefined{shadecolor}{\definecolor{shadecolor}{rgb}{.97, .97, .97}}
\makeatother
\ifLuaTeX
  \usepackage{selnolig}  
\fi
\usepackage[]{natbib}
\IfFileExists{bookmark.sty}{\usepackage{bookmark}}{\usepackage{hyperref}}
\IfFileExists{xurl.sty}{\usepackage{xurl}}{} 
\hypersetup{
  pdftitle={Editorial: Guidelines and Best Practices to Share Deidentified Data and Code},
  pdfauthor={Nicholas Jon Horton; Sara Stoudt},
  pdfkeywords={no keywords for editorial},
  colorlinks=true,
  linkcolor={blue},
  filecolor={Maroon},
  citecolor={Blue},
  urlcolor={Blue},
  pdfcreator={LaTeX via pandoc}}

\begin{document}

\def\spacingset#1{\renewcommand{\baselinestretch}%
{#1}\small\normalsize} \spacingset{1}


\date{July 8, 2024}
\title{\bf Editorial: Guidelines and Best Practices to Share
Deidentified Data and Code}
\author{
Nicholas Jon Horton\\
Department of Mathematics and Statistics, Amherst College\\
and\\Sara Stoudt\\
Department of Mathematics, Bucknell University\\
}
\maketitle

\bigskip
\bigskip
\begin{abstract}
No abstract for editorial
\end{abstract}

\noindent%
\vfill

\newpage
\spacingset{1.9} 
\ifdefined\Shaded\renewenvironment{Shaded}{\begin{tcolorbox}[frame hidden, interior hidden, borderline west={3pt}{0pt}{shadecolor}, sharp corners, boxrule=0pt, enhanced, breakable]}{\end{tcolorbox}}\fi

\hypertarget{introduction}{%
\section{Introduction}\label{introduction}}

In 2022, the \emph{Journal of Statistics and Data Science Education}
(JSDSE) instituted augmented requirements for authors to post
deidentified data and code underlying their papers
\citep{jsdse_reproducibility:2022}. These changes were prompted by an
increased focus on reproducibility and open science \citep{nasem2019}. A
recent review of data availability practices noted that ``such policies
help increase the reproducibility of the published literature, as well
as make a larger body of data available for reuse and re-analysis''
\citep{plos_one}.

\emph{JSDSE} values accessibility as it endeavors to share knowledge
that can improve educational approaches to teaching statistics and data
science. Because institution, environment, and students differ across
readers of the journal, it is especially important to facilitate the
transfer of a journal article's findings to new contexts. This process
may require digging into more of the details, including the deidentified
data and code used to support the article's findings.

Two years on, we felt that it was valuable to provide our readers and
authors with a review of why the requirements for code and data sharing
were instituted, summarize ongoing trends and developments in open
science, discuss options for data and code sharing, and share advice for
authors.

\hypertarget{why-is-data-and-code-sharing-necessary}{%
\section{Why is data and code sharing
necessary?}\label{why-is-data-and-code-sharing-necessary}}

Why the push for the sharing of data and code? Data sharing, as well as
the sharing of the code for the processing and analysis steps, is a
necessity if readers or other researchers are to reproduce findings
published in the journal. Beyond building trust in a particular
article's findings, the sharing of deidentified data can help others in
their own studies.

An editorial in \emph{PLOS ONE} \citeyearpar{plos_data} states that:

``Data availability allows and facilitates:

\begin{enumerate}
\def\labelenumi{\arabic{enumi}.}
\tightlist
\item
  Validation, replication, reanalysis, new analysis, reinterpretation or
  inclusion into meta-analyses;
\item
  Reproducibility of research;
\item
  Efforts to ensure data are archived, increasing the value of the
  investment made in funding scientific research;
\item
  Reduction of the burden on authors in preserving and finding old data,
  and managing data access requests;
\item
  Citation and linking of research data and their associated articles,
  enhancing visibility and ensuring recognition for authors, data
  producers and curators.''
\end{enumerate}

Funding bodies have been requiring researchers to share their data as
well. In August 2022, the White House Office of Science and Technology
Policy directed federal agencies to update their data-sharing policies
such that all federally-funded research data would have to be shared
freely by 2025 \citep{dubois:2023}. The National Institute of Health's
Generalist Repository Ecosystem Initiative has taken a comprehensive
approach to foster data and workflow sharing \citep{GREI}. A special
theme on ``Changing the Culture on Data Management and Data Sharing in
Biomedicine'' was published in 2022 \citep{hdsr_theme}.

But it's not just deidentified data that are needed. As but one example
of the importance of code sharing, \citet{nature_246} found that when a
large group of biologists analyzed the same dataset, their results were
widely divergent.

A NASEM report on \emph{Reproducibility and Replicability in Science}
concluded that ``Journal editors should consider ways to ensure
reproducibility for publications that make claims based on computations,
to the extent ethically and legally possible (page 2)''
\citep{nasem2019}.

A recent editorial in the Journal of the American Statistical
Association (JASA), the flagship journal of the American Statistical
Association (ASA) noted that their reproducibility process ``fosters a
culture of transparency and accountability that is critical to nurture
the trust placed by the lay public in scientific research''
\citep{jasa:2024}.

We anticipate that other entities will be moving in this direction in
the future.

\hypertarget{how-have-other-journals-responded}{%
\section{How have other journals
responded?}\label{how-have-other-journals-responded}}

In the past ten years, many top journals have added requirements to
support reproducible research, including policies on data sharing. In
March 2014, \emph{PLOS ONE} started requiring that the supporting data
for the results in its publications be shared \citep{plos_one}. The
\emph{Harvard Data Science Review} released a data sharing requirement
in 2021 \citep{hdsr}.

In September 2016, \emph{Nature} started to require that data
availability statements be included as part of its publications.
Although this policy doesn't require that the data actually be made
available, the intention of the data availability statement is to make
the authors' decision to share or not share the data more transparent.

The ASA released recommendations in 2018 for their journals that are
jointly published with Taylor \& Francis (this set of journals includes
\emph{JSDSE}). These recommendations encourage data sharing and the
inclusion of data availability statements \citep{asa_policy}.

In 2016, \emph{JASA} began to require both code and data at the revision
stage of submissions to the ``Applications and Case Studies'' section
\citep{jasa_repro}. \emph{JASA} editors extended this requirement to the
revision stage of submissions to the ``Theory and Methods'' section in
2021.

\hypertarget{how-best-to-share-deidentified-data}{%
\section{How best to share deidentified
data}\label{how-best-to-share-deidentified-data}}

\hypertarget{what-datasets-can-be-shared}{%
\subsection{What datasets can be
shared?}\label{what-datasets-can-be-shared}}

Many journals' data sharing requirements reference a ``minimum'' or
``minimal'' dataset that is required to be shared. For example,
\emph{PLOS ONE} considers a minimal dataset as one that has ``the data
required to replicate all study findings reported in the article, as
well as related metadata and methods'' \citep{plos_data} while
\emph{Nature}'s minimum dataset is one that is ``necessary to interpret,
verify and extend the research in the article, transparent to readers''
\citep{nature_portfolio}. The FAIR Data Principles of data sharing
request that these shared datasets are Findable, Accessible,
Interoperable , and Reusable with the goal of increasing ``knowledge
discovery and innovation'' \citep{fair}.

At times, investigators have gone back to their Institutional Review
Board (IRB) to seek approval for requests to retroactively share
deidentified data. At the heart of such a request is the creation of a
``minimal'' dataset without any direct or indirect identifiers.

As one example, this process was undertaken to make the data from the
Health Evaluation and Linkage to Primary Care (HELP) study
\citep{same:lars:hort:2003} available on the Center for Open Science's
Open Science Framework (OSF) \citep{HELP}.

However, sharing a ``minimum'' dataset that just elides identifiers may
not be sufficient to avoid reidentification. When creating a ``minimum''
dataset it may be necessary to only share a proper subset of the
attributes recorded on individual subjects to minimize risk of
de-anonymization.

Moving forward, researchers need to be thinking about data sharing when
they begin to design their studies. The DMP Tool offers useful guidance
for investigators designing data management and sharing plans
\citep{dmptool}.

Other approaches may be needed, depending on the study. Consider a
hypothetical qualitative research example where students are being
interviewed before graduating to find out about their trajectory through
a data science program. They are asked a series of questions, and their
responses are transcribed. The full text of the interview would reveal a
lot of personally identifiable information. However, the transcripts are
often coded by researchers as a pre-processing step before analysis.
These codes are what are used in the analysis phase of the project, so
having access to the coded data rather than the raw transcript data
would be enough to reproduce the work. \emph{JSDSE} has published papers
involving the analysis of qualitative data \citep{lesser:2024} where the
coded dataset is shared (and not the full transcripts, video recordings,
etc.). \citet{dubois:2023} considers a similar question.

Many journals' data sharing requirements provide caveats and exceptions
for situations where data sharing may be problematic, especially with
respect to privacy, and leave room for more controlled access. Similar
policies operate for \emph{JSDSE} (see Section~\ref{sec-waiver}).

\hypertarget{what-file-formats-and-licensing}{%
\subsection{What file formats and
licensing?}\label{what-file-formats-and-licensing}}

Making data ``interoperable,'' as the FAIR Data Principles advocate for,
includes sharing it in a non-proprietary file format (such as CSV or
TXT) so that it is easily accessed across platforms and using a variety
of software. Making data ``reusable'' includes sharing it with a clear
usage license for those who want to use it in their own work. Recent
papers published in \emph{JSDSE} are made available under a Creative
Commons license, with copyright held by the author. This gives readers
the right to build upon the work. The ``accessible'' piece of FAIR Data
principles speaks to making data freely available, which can broaden who
can do further research on the topic \citep{nagaraj:2020}.

\hypertarget{which-repository-to-use}{%
\subsection{Which repository to use?}\label{which-repository-to-use}}

Journals like \emph{Nature} \citep{natureP} and \emph{Proceedings of the
National Academy of Sciences} (PNAS) \citep{pnasP} provide detailed
guidelines on appropriate repositories for sharing data; these can be
used as references for exploring repository options.

The FAIR Data Principles also provide a framework for determining what
makes a ``good'' repository. For example, part of the ``findable'' piece
of the FAIR Data Principles involves having a way to link to the data
that is permanent. A digital object identifier (DOI) is an example of a
permanent, FAIR, way to access a dataset. A GitHub repository is not
``findable'' in this sense because the repository can be deleted,
renamed, or reorganized, such that the link to the repository no longer
brings a reader to a dataset as intended \citep{fairdata}.

\emph{JSDSE} authors have deposited data in a variety of FAIR
repositories. A number have utilized the OSF \citep{alzen:2024} which
facilitates anonymization of authors (that can then be unmasked once a
paper has been accepted) \citep{osfguide}. Others have used Zenodo
\citep{mocko:2024} and Mendeley Data \citep{miller:2024}, sometimes
based on recommendations or requirements of their institutions.

\hypertarget{challenges-to-sharing-of-deidentified-data}{%
\section{Challenges to sharing of deidentified
data}\label{challenges-to-sharing-of-deidentified-data}}

Although broad access is a value, the devil really is in the details.
There is a need to balance both transparency and reproducibility with
data privacy \citep{zasl:hort:1998}. Many ethical considerations exist
for data that often comes from students.

While many investigations might be deemed exempt from human subjects
oversight (involving no or minimal risk), we don't have to look far to
come up with a plausible, yet, more complicated scenario that our
journal has already faced.

Suppose each row in a dataset is a student. There are columns for
performance on a pre-test, performance on a post-test, and a variety of
demographic data including race, major, gender, and class year. If this
data were shared without any names of the students, that would on the
surface look deidentified. However, the combination of demographic
variables can reveal individuals, especially those belonging to
intersectional, minority groups. For example, if there was only one
first-year, Asian, male student in the class, we automatically know how
they performed on the pre-test and post-test, which violates this
student's privacy. (Even if there were more than one such student, the
probability of reidentification is non-zero.)

One option is to remove the sensitive demographic columns of the data,
and focus on the response and covariates of interest. If we are
interested in the relationship between pre-test and post-test scores,
the demographic variables are not essential to reproduce the analysis.
Removing potentially identifying demographic variables in this way and
sharing a deidentified dataset that is a subset of the full data is the
approach taken by \citet{evans:2023}.

However, if those demographic variables are part of the research
question, i.e., we want to investigate how demographic variables may
mediate the relationship between pre-test and post-test performance,
then these variables would be an important part of the reproducibility
process. Another option might be to create synthetic data
\citep{synthetic} that has similar characteristics to the raw data but
where each row does not represent an actual student.

The idea here is that the shared processing and analysis code could be
run using the synthetic data, but no student data would be revealed.
Such an approach would be indicated if there was substantial risk of
disclosure (typically beyond what is considered ``exempt from human
subjects oversight'' by many IRBs).

\hypertarget{answers-to-frequently-asked-questions}{%
\section{Answers to Frequently Asked
Questions}\label{answers-to-frequently-asked-questions}}

\hypertarget{what-should-a-data-availability-statement-look-like}{%
\subsection{What should a data availability statement look
like?}\label{what-should-a-data-availability-statement-look-like}}

Even in a simple scenario, we can see that each dataset comes with its
own nuance. Therefore, in answer to this question, we resort to a
statistician's or data scientist's favorite phrase: it depends!

Articles published in \emph{JSDSE} are not intended to adhere to a
single, rigid, data availability policy but rather the intent of
requiring a data availability statement at all is to start a
conversation and a creative problem solving process. The journal wants
to work with authors to negotiate a minimal dataset that balances the
dueling goals of accessibility and protecting the rights of the data
subjects.

That being said, we provide several examples of possible data
availability statements that meet the letter and spirit of the
guidelines:

\begin{enumerate}
\def\labelenumi{\arabic{enumi}.}
\item
  The deidentified data and code that support the findings of this study
  are openly available at the Open Science Framework (OSF): provide URL
  here.
\item
  The authors confirm that the data supporting the findings of this
  study are available within the article.
\item
  Data sharing is not applicable to this article as no new data were
  created or analyzed in this study.
\end{enumerate}

Student data can come in many forms, their responses to a survey, their
performance on an assessment, their demographics, but they can also come
in a variety of types including qualitative data. Written responses,
audio/video, and interview transcripts are personal and can contain more
identifiable information than quantitative data, which can make sharing
the raw data challenging. Such a study might adopt the following data
availability statement:

\begin{enumerate}
\def\labelenumi{\arabic{enumi}.}
\setcounter{enumi}{3}
\tightlist
\item
  Due to the nature of the research, the video recordings of the
  participants are not available. The deidentified coded data and
  analysis code that support the findings of this study are openly
  available at the Open Science Framework (OSF): provide URL here.
\end{enumerate}

\hypertarget{is-share-upon-request-an-acceptable-option}{%
\subsection{Is share upon request an acceptable
option?}\label{is-share-upon-request-an-acceptable-option}}

Alternative approaches to data availability statements such as ``share
upon request'' have been notoriously ineffective as a way of fostering
data and code sharing \citep{tedersoo:2021, gabelica}. ``Share upon
request'' is not, in general, allowed for \emph{JSDSE} papers.

\hypertarget{sec-waiver}{%
\subsection{How can a waiver be requested?}\label{sec-waiver}}

Authors seeking an exemption to deidentified data sharing should send a
formal request to the editor asking for a waiver. Such a request should
include (1) the justification for why a deidentified dataset cannot be
made available, (2) background on the study including the original IRB
protocol and study documents, and (3) a codebook for the data included
in the waiver.

\hypertarget{where-should-the-data-availability-statement-be-located}{%
\subsection{Where should the data availability statement be
located?}\label{where-should-the-data-availability-statement-be-located}}

In terms of placement, the data availability statement should appear
just before the references. Information about where deidentified data
and code can be found is also provided via metadata input during the
submission process. A stub data availability statement is included in
the template (LaTeX and Quarto) provided for authors \citep{template}.

\hypertarget{where-can-i-learn-more-about-computational-reproducibility}{%
\subsection{Where can I learn more about computational
reproducibility?}\label{where-can-i-learn-more-about-computational-reproducibility}}

This editorial has focused most heavily on the details surrounding the
sharing of deidentified data as privacy concerns make additional
guidance necessary. However, the sharing of the code used to process and
analyze that data is also a necessary but insufficient step to foster
computational reproducibility \citep{tier}. \citet{Sandve:2013} provide
ten simple rules for reproducible computational research. While all are
important, rule ten describes ways to provide public access to scripts
and results. \citet{Ball:2022} describe approaches for teaching
reproducible science. We encourage authors to take advantage of the
Quarto paper template \citep{quarto} that has been made available to
facilitate reproducible analysis and reporting.

\hypertarget{closing-thoughts}{%
\section{Closing thoughts}\label{closing-thoughts}}

We realize that best practices to foster reproducibility and
replicability are fast changing and require additional efforts by
authors, reviewers, and editors.

This process will require updates to our standard procedures, including
improved templates for human subjects oversight. For example, protocols
will need to no longer state that ``deidentified data will be destroyed
at the end of the study''. It will be equally important to inform
subjects that fully deidentified data will be made available and not
just summarized in aggregate.

However, we believe that these changes can and should be undertaken, and
that these changes will allow us as a community to balance privacy and
sharing in a way that fosters better science.

At present, the deidentified data and code included with submissions are
not formally included in the review process (though some reviewers and
associate editors may choose to incorporate this information). In the
future, the journal may consider formalizing this review and/or starting
up a reproducibility process such as that described by \citet{jasa:2024}
or the creative approach undertaken by the journals published by the
American Economic Association \citep{vilh:2022}.

We hope that this discussion will benefit future authors and readers as
they navigate this new territory. We concur with the assessment of
\citet{jasa:2024} which stated that ``the journal gains credibility from
a system that holds researchers accountable for their work, credibility
that can be leveraged to publish controversial and impactful research
that has the potential to change a field of study''.

Nicholas J. Horton ORCID: https://orcid.org/0000-0003-3332-4311

Sara Stoudt ORCID: https://orcid.org/0000-0002-1693-8058

\hypertarget{acknowledgements}{%
\section{Acknowledgements}\label{acknowledgements}}

We acknowledge helpful comments on an earlier draft of the manuscript
from Richard Ball, Mine Çetinkaya-Rundel, Johanna Hardin, Chris
Paciorek, Joshua Rosenberg, and Jeffrey Witmer.

\hypertarget{disclosure-statement}{%
\section*{Disclosure statement}\label{disclosure-statement}}
\addcontentsline{toc}{section}{Disclosure statement}

The authors have no conflicts of interest to declare.

\hypertarget{data-availability-statement}{%
\section*{Data availability
statement}\label{data-availability-statement}}
\addcontentsline{toc}{section}{Data availability statement}

No data or code are associated with this editorial.

  \bibliography{bibliography.bib}

\end{document}